\documentclass[prd,aps,preprintnumbers,floatfix,showpacs,preprint,tightenlines,superscriptaddress]{revtex4}
\usepackage{epsfig}
\usepackage{graphics}
\usepackage{latexsym}
\usepackage{amsmath}
\usepackage{amssymb}
\usepackage{rotating}

\newcommand{\gev}{\,{\rm GeV}}
\newcommand{\fm}{\,{\rm fm}}
\newcommand{\mpi}{m_\pi}
\newcommand{\chpt}{$\chi$PT}

\begin{document}

\preprint{JLAB-THY-08-886, ADP-08-08-T668, ANL-PHY-12172-TH-2008}

\title{Chiral extrapolation of octet-baryon charge radii\footnote
{Notice: Authored by Jefferson Science Associates, LLC under U.S.
DOE Contract No. DE-AC05-06OR23177. The U.S. Government retains a
non-exclusive, paid-up, irrevocable, world-wide license to publish
or reproduce this manuscript for U.S. Government purposes.}}
\author{P. Wang}
\affiliation{Jefferson Laboratory, 12000 Jefferson Ave., Newport
News, VA 23606 USA} \affiliation{Theoretical Physics Center for
Science Facilities (TPCSF), CAS, P. R. China}

\author{D. B. Leinweber}
\affiliation{Special Research Center for the Subatomic
Structure of Matter (CSSM) and Department of Physics, University of
Adelaide 5005, Australia}

\author{A. W. Thomas}
\affiliation{Jefferson Laboratory, 12000 Jefferson Ave., Newport
News, VA 23606 USA}
\affiliation{Department of Physics, College of
William and Mary, Williamsburg VA 23187 USA}

\author{R. D. Young}
\affiliation{Physics Division, Argonne National Laboratory,
Argonne, IL 60439, USA}

\begin{abstract}

  The charge radii of octet-baryons obtained in quenched lattice-QCD
  calculations are extrapolated within heavy-baryon chiral
  perturbation theory.  Finite-range regularization (FRR) is applied
  to improve the convergence of the chiral expansion and to provide
  estimates of quenching artifacts.  Lattice values of quark
  distribution radii and baryon
  charge radii for $\mpi^2$ in the range (0.1, 0.7)$\gev^2$ are described very well
  with FRR.  Upon estimating corrections for both finite-volume and
  quenching effects, the obtained charge radii of the proton, neutron
  and $\Sigma^-$ are in good agreement with experimental
  measurements. The predicted charge radii of the remaining
  octet-baryons
  have not yet been measured and present a challenge to future
  experiments.

\end{abstract}




\maketitle

\section{Introduction}

The study of the electromagnetic form factors of baryons is of
crucial importance to understanding the non-perturbative properties
of QCD. Though QCD is accepted as the fundamental theory of the
strong interaction, it remains a theoretical challenge to
quantitatively probe the non-perturbative domain. There are many
effective methods and phenomenological models which have been
applied to study the electromagnetic properties of baryons: the
cloudy bag model \cite{Lu:1997sd}, the constituent quark model
\cite{Berger:2004yi,JuliaDiaz:2003gq}, the $1/N_c$ expansion
approach \cite{Buchmann:2002et}, the perturbative chiral quark model
\cite{Cheedket:2002ik}, the extended vector meson dominance model
\cite{Williams:1996id}, the quark-diquark model
\cite{Hellstern:1995ri} and the Schwinger-Dyson formalism
\cite{Oettel:1998bk,Alkofer:2004yf,Eichmann:2007nn}. Various
formulations of heavy-baryon chiral perturbation theory (\chpt) have
also been widely applied to this problem
\cite{Puglia:2000jy,Fuchs:2003ir,Kubis:2000aa,Kubis:2000zd}. It has
been observed that expansions in \chpt\ are consistent with
experimental results up to $Q^2 \simeq 0.1\gev^2$
\cite{Fuchs:2003ir}. Extensions of \chpt\ to explicitly incorporate
vector mesons have been demonstrated to improve the applicability to
$Q^2 \simeq 0.4\gev^2$ \cite{Schindler:2005ke}.

As well as the above model calculations, the past few years have
seen increased activity in lattice-QCD studies of the
electromagnetic form factors. Significant efforts to probe baryon
electromagnetic structure in lattice QCD have been driven by the
Adelaide group \cite{Zanotti:2003gc,Boinepalli:2006xd}, the Cyprus
group \cite{Alexandrou:2006ru}, and the QCDSF
\cite{Gockeler:2003ay,Gockeler:2006ui,Gockeler:2007hj} and LHP
Collaborations \cite{Edwards:2005kw,Alexandrou:2005fh}. While
lattice-QCD provides the strongest tool for studying
non-perturbative phenomena in QCD, it does come with its own
challenges. In particular, artifacts of (unitary) lattice
simulations arising from discretisation, finite-volume and
unphysical quark masses all need to be carefully dealt with to
extract the predictions of QCD relevant to the real world. (There
are further, more complicated issues in dealing with non-unitary
approximations: partial- and full-quenching, 4th-rooting,
mixed-action etc.).

The principle focus of this manuscript is on the quark-mass
dependence of baryon charge radii and the chiral extrapolation of
lattice simulation results performed at unphysically large quark
masses. Characterizing the quark-mass dependence of hadronic
observables in QCD can be achieved within the low-energy effective
theory of QCD, \chpt --- see for example, Bernard's recent review on
baryon phenomena \cite{Bernard:2007zu}.

A celebrated feature of such effective theories is the
model-independence of leading nonanalytic contributions to
quark-mass expansions \cite{Li:1971vr}. Such {\em chiral logs} are a
direct consequence of the spontaneously broken chiral symmetry of
QCD. Neglecting such behavior in chiral extrapolations (of even
high-quality lattice calculations) can potentially render results
which have less resemblance of QCD than some of the models discussed
above. While the inclusion of such logs in chiral extrapolations is
necessary to maintain QCD in the extraction of physical results
\cite{Leinweber:1992hj,Leinweber:1998ej,Leinweber:2001ui,Detmold:2001jb,Hemmert:2002uh},
this can be a challenging task because of the poor convergence
properties of the EFT expansion at moderate quark masses
\cite{Young:2002ib,Young:2004tb,Beane:2004ks,Djukanovic:2006xc}.

In this manuscript we work with finite-range regularisation (FRR)
to improve the convergence properties of the quark-mass
expansion of the EFT \cite{Young:2002ib,Young:2004tb,Leinweber:2003dg}.

As an effective theory describing long-distance phenomena, \chpt\ can
also describe the finite-volume effects of restricting the underlying
theory to finite boundary conditions \cite{Gasser:1987zq}. The
accuracy of these corrections is also subject to the usual conditions,
being at light-enough quark masses and large-enough volumes. In this
work we impose the finite boundary conditions on the relevant one-loop
graphs. These provide the leading estimates of the finite-volume
corrections.

The lattice simulation results studied in this paper come from the
CSSM Lattice Collaboration \cite{Boinepalli:2006xd}. These lattice
results have been evaluated with quenched gauge-field ensembles. With
the underlying dynamics modified by neglecting the quark loops of the
QCD vacuum, the effective low-energy theory is described by
quenched \chpt\ (Q\chpt). For baryons, this was first formulated by
Labrenz and Sharpe \cite{Labrenz:1996jy} and for relevant work on the
electromagnetic form factors in quenched and partially-quenched
theories see
Refs.~\cite{Savage:2001dy,Leinweber:2002qb,Arndt:2003ww,Tiburzi:2004mv}.

Given quenched lattice results, we use the leading meson-loop diagrams
to estimate the corrections in going to the fully dynamical theory
\cite{Young:2004tb}. This is based on the empirical observation that
the discrepancies between the quenched and dynamical nucleon and
Delta-baryon masses are well described by the associated leading
meson-loop dressings, as evaluated with a dipole finite-range
regulator \cite{Young:2002cj}. The physical picture drawn from these
results is quite intuitive, once the interquark forces are matched at
an intermediate distance scale (in this case the Sommer scale
\cite{Edwards:1997xf}) the differing long-range features are
dominantly described by the low-energy EFT. In this case the FRR scale
acts to separate the long from the short. Further, any residual
difference in the $q\bar{q}$ force at quite short distances does not
appear to play a significant role in the bulk, low-energy structure.

In section II, we will briefly introduce the chiral Lagrangian which
is used for the octet charge form factors. Charge form factors and
radii are calculated in section III with quenched and full QCD.
Numerical results are presented in section IV and finally section V
is the summary.

\section{Chiral Lagrangian}

There are many papers which deal with heavy baryon chiral
perturbation theory. For details see for example, Refs.
\cite{Jenkins:1990jv,Bernard:1992qa,Bernard:2007zu}. For
completeness, we briefly introduce the formalism in this section. In
heavy baryon chiral perturbation theory, the lowest chiral
Lagrangian for the baryon-meson interaction which will be used in
the calculation of the octet-baryon charge form factors is
\begin{eqnarray}
{\cal L}_v &=&i{\rm Tr}\bar{B}_v(v\cdot {\cal D}) B_v+2D{\rm
Tr}\bar{B}_v S_v^\mu\{A_\mu,B_v\} +2F{\rm Tr}\bar{B}_v
S_v^\mu[A_\mu,B_v]
\nonumber \\
&& -i\bar{T}_v^\mu(v\cdot {\cal D})T_{v\mu} +{\cal C}(\bar{T}_v^\mu
A_\mu B_v+\bar{B}_v A_\mu T_v^\mu),
\end{eqnarray}
where $S_\mu$ is the covariant spin-operator defined as
\begin{equation}
S_v^\mu=\frac i2\gamma^5\sigma^{\mu\nu}v_\nu.
\end{equation}
Here, $v^\nu$ is the nucleon four velocity (in the rest frame, we
have $v^\nu=(1,\vec{0})$). We incorporate the explicit propagation
of octet and decuplet baryon states, with $D$, $F$ and $\cal C$
denoting the relevant meson-baryon couplings (which, in principle,
are to be determined in the chiral limit).  The chiral covariant
derivative $D_\mu$ is written as $D_\mu B_v=\partial_\mu
B_v+[V_\mu,B_v]$. The pseudoscalar meson octet couples to the baryon
field through the vector and axial vector combinations
\begin{equation}
V_\mu=\frac12(\zeta\partial_\mu\zeta^\dag+\zeta^\dag\partial_\mu\zeta),\qquad
A_\mu=\frac12(\zeta\partial_\mu\zeta^\dag-\zeta^\dag\partial_\mu\zeta),
\end{equation}
where
\begin{equation}
\zeta=e^{i\phi/f}, \qquad f=93~{\rm MeV}.
\end{equation}
The matrix of pseudoscalar fields $\phi$ is expressed as
\begin{eqnarray}
\phi=\frac1{\sqrt{2}}\left(
\begin{array}{lcr}
\frac1{\sqrt{2}}\pi^0+\frac1{\sqrt{6}}\eta & \pi^+ & K^+ \\
\pi^- & -\frac1{\sqrt{2}}\pi^0+\frac1{\sqrt{6}}\eta & K^0 \\
K^- & \bar{K}^0 & -\frac2{\sqrt{6}}\eta
\end{array}
\right).
\end{eqnarray}
$B_v$ and $T^\mu_v$ are velocity-dependent fields which are
related to the original baryon octet and decuplet fields $B$ and
$T^\mu$ by
\begin{equation}
B_v(x)=e^{im_N \not v v_\mu x^\mu} B(x),
\end{equation}
\begin{equation}
T^\mu_v(x)=e^{im_N \not v v_\mu x^\mu} T^\mu(x).
\end{equation}
In the chiral $SU(3)$ limit, the octet baryons are degenerate.  As
the physical strange-quark mass is significant, we incorporate the
physical mass-splittings in the evaluation of the chiral-loop
diagrams.

The explicit form of the octet-baryon matrix is written as
\begin{eqnarray}
B=\left(
\begin{array}{lcr}
\frac1{\sqrt{2}}\Sigma^0+\frac1{\sqrt{6}}\Lambda &
\Sigma^+ & p \\
\Sigma^- & -\frac1{\sqrt{2}}\Sigma^0+\frac1{\sqrt{6}}\Lambda & n \\
\Xi^- & \Xi^0 & -\frac2{\sqrt{6}}\Lambda
\end{array}
\right).
\end{eqnarray}
The baryon decuplets are defined by the rank-3 symmetric tensor, with
unique elements given by
\begin{eqnarray}
&&T_{111}=\Delta^{++}, \quad T_{112}=\frac1{\sqrt{3}}\Delta^+, \quad
T_{122}=\frac1{\sqrt{3}}\Delta^0, \quad T_{222}=\Delta^-, \nonumber\\
&&T_{113}=\frac1{\sqrt{3}}\Sigma^{\ast,+}, \quad
T_{123}=\frac1{\sqrt{6}}\Sigma^{\ast,0}, \quad
T_{223}=\frac1{\sqrt{3}}\Sigma^{\ast,-}, \nonumber \\
&&T_{133}=\frac1{\sqrt{3}}\Xi^{\ast,0}, \quad
T_{233}=\frac1{\sqrt{3}}\Xi^{\ast,-}, \quad T_{333}=\Omega^{-}.
\end{eqnarray}

In the heavy-baryon formalism, the propagators of the octet and
decuplet baryon, $j$, are respectively expressed as
\begin{equation}
\frac i {v\cdot k-\Delta^{jB}+i\varepsilon} ~~{\rm and}~~ \frac
{iP^{\mu\nu}} {v\cdot k-\Delta^{jB}+i\varepsilon},
\end{equation}
with $P^{\mu\nu}=v^\mu v^\nu-g^{\mu\nu}-(4/3)S_v^\mu S_v^\nu$.
$\Delta^{ab}=m_b-m_a$ is the mass difference of between the two
baryons. The propagator of meson $k$ ($k=\pi$, $K$, $\eta$) is the
usual free propagator, i.e.:
\begin{equation}
\frac i {k^2-m_k^2+i\varepsilon}.
\end{equation}

\section{charge form factors and radii}
In the heavy-baryon formalism, the baryon form factors are defined
as
\begin{equation}
<B(p^\prime)|J_\mu|B(p)>=\bar{u}(p^\prime)\left\{v_\mu
G_E(Q^2)+\frac{i\epsilon_{\mu\nu\alpha\beta}v^\alpha S_v^\beta
q^\nu}{m_N}G_M(Q^2)\right\}u(p),
\end{equation}
where $q=p^\prime-p$ and $Q^2=-q^2$. In this paper, we focus on the
charge form factors since we will extrapolate octet-baryon charge
radii.

\begin{center}
\begin{figure}[hbt]
\includegraphics[scale=0.5]{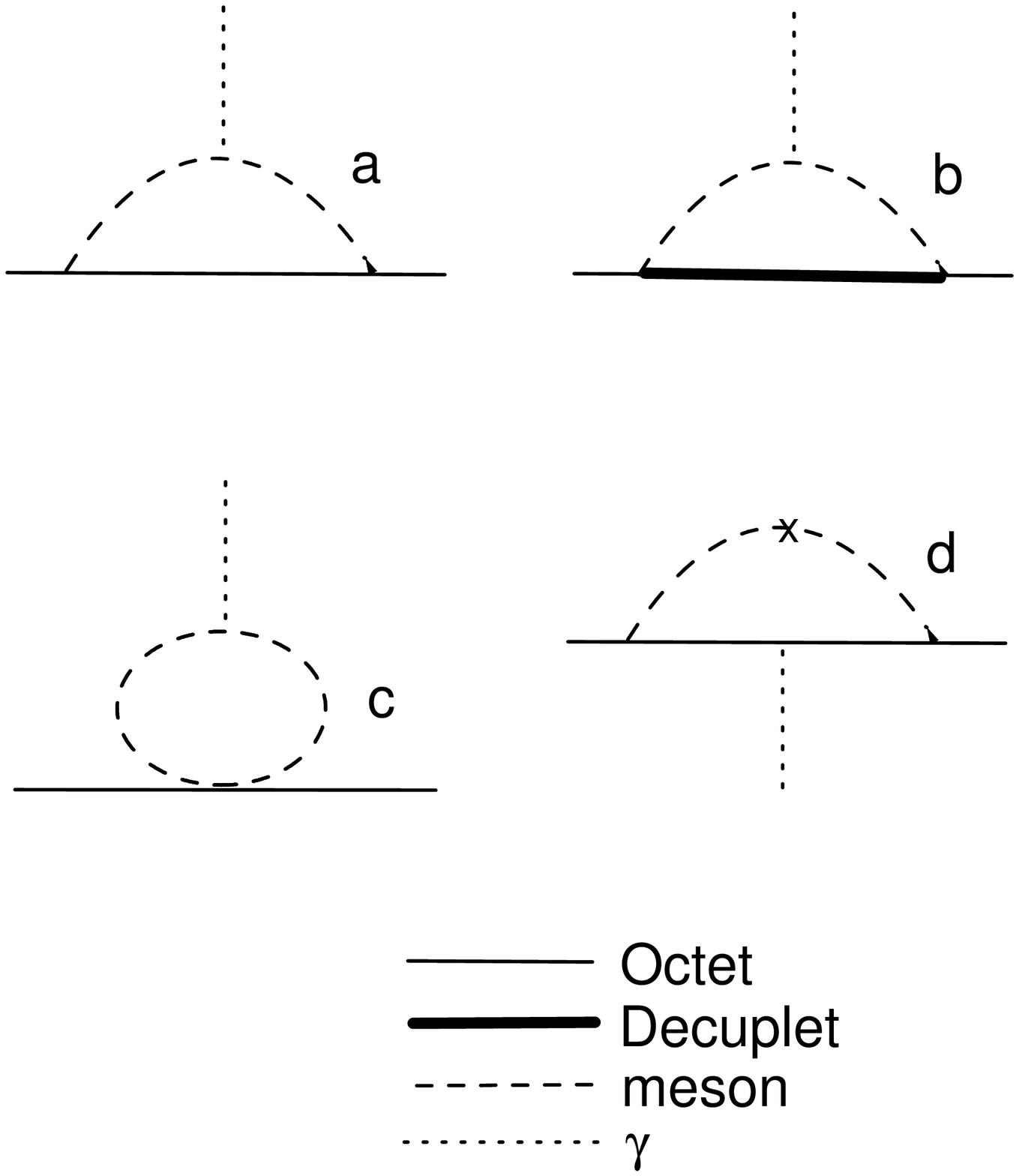}
\caption{Leading and next to leading order diagrams for the baryon
charge form factors. Diagram d contributes only in the quenched
case. \label{fig:diag}}
\end{figure}
\end{center}

With the Lagrangian, the diagrams for the charge form factors are
shown in Fig.~\ref{fig:diag}. Our counting is ordered in terms of
powers of the quark mass contributing to the given electric radii.
For the purposes of counting, where the typical baryon mass
splitting is small relative to the lattice meson masses, we treat
the power of the quark mass that defines the order of a given
diagram by neglecting the mass splitting, $\Delta$. The diagrams are
nevertheless evaluated with the splittings restored. In the limit
$m_\pi << \Delta$, diagrams a with the intermediate nucleon state
and c give rise to the leading-log divergence of charge radii in
full QCD. Diagram b gives a next to leading order nonanalytic term
associated with the mass difference between octet and decuplet
baryons. For the intermediate hyperson states of diagram a, the mass
difference between hyperons and nucleon also provides a next to
leading order nonanalytic term as diagram b. In this limit, we only
have a partial contribution of next order. This is certainly one of
the places where our analysis is somewhat phenomenological. In the
quenched case, each diagram has a different coefficient from that of
dynamical QCD. In particular, diagram c has no contribution in the
quenched case. Further, the double hairpin diagram d contributes
only in the quenched case, where the $\eta^{'}$ is degenerate with
the pion.

\begin{center}
\begin{figure}[hbt]
\includegraphics[scale=0.5]{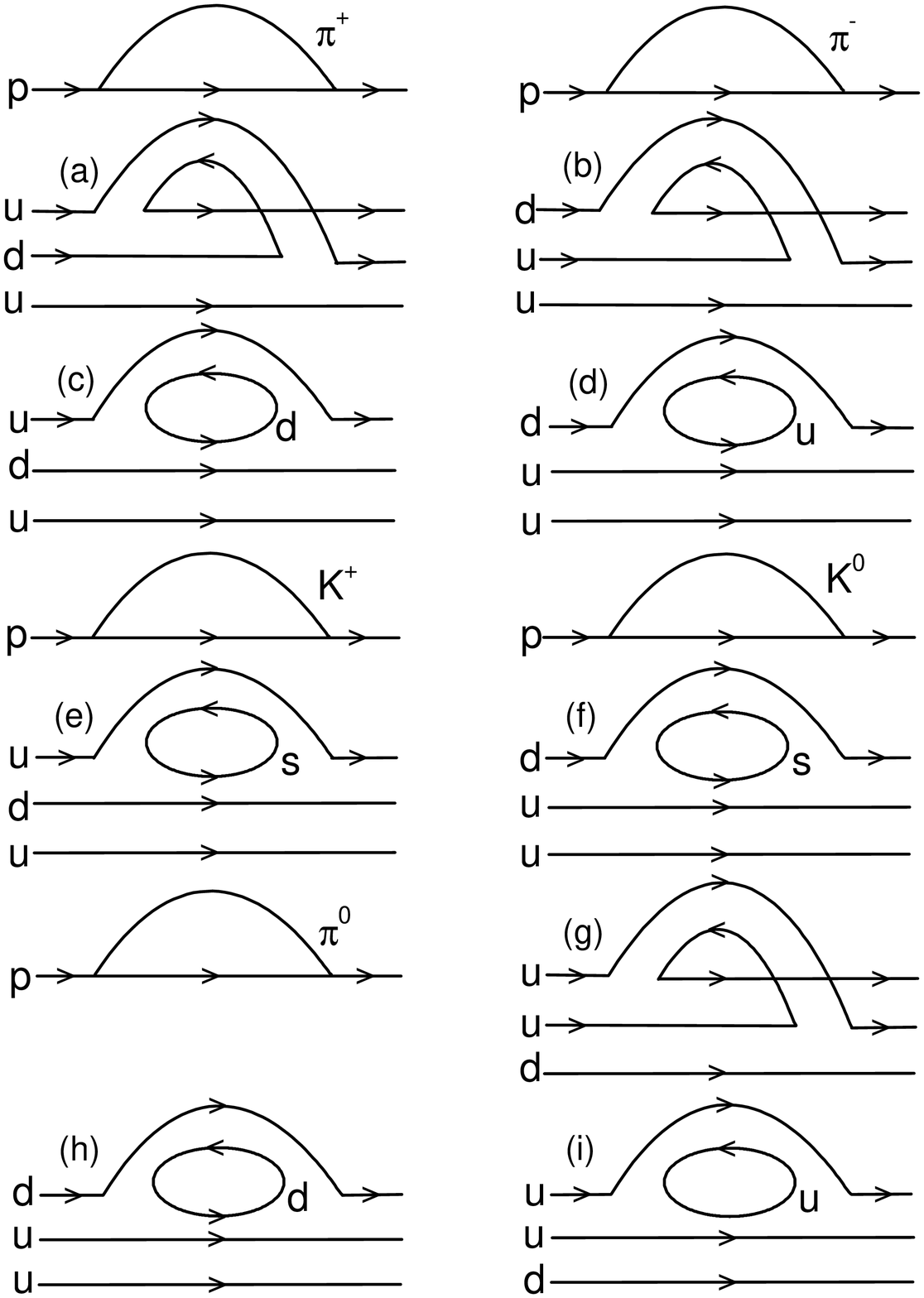}
\caption{Feymann diagrams of Fig.~1a for the proton, in terms of
quark lines. \label{fig:quench}}
\end{figure}
\end{center}

\begin{center}
\begin{figure}[hbt]
\includegraphics[scale=0.5]{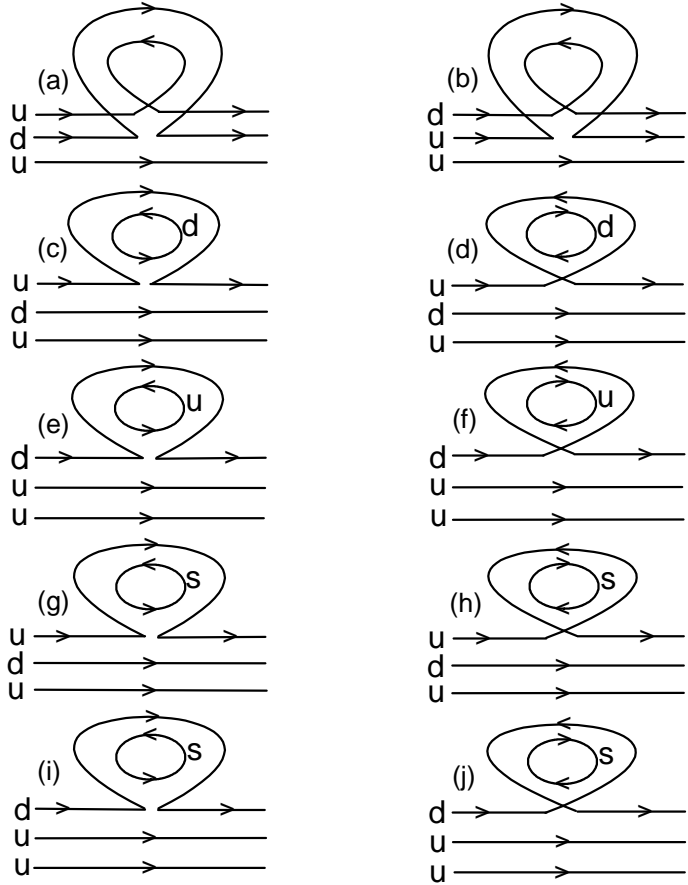}
\caption{Feymann diagrams of Fig.~1c for the proton, in terms of
quark lines. \label{fig:quench2}}
\end{figure}
\end{center}

Upon integration over $k_0$, the contribution to charge form factors
of Fig.~1a can be expressed as
\begin{equation}\label{gea}
G_E^{(a)}(Q^2)=\frac{\beta_E^a}{16\pi^3f_\pi^2}\int
d\vec{k}\frac{
u(\vec{k})u(\vec{k}-\vec{q})
\vec{k}\cdot (\vec{k}-\vec{q})}
{\omega(\vec{k})\omega(\vec{k}-\vec{q})
(\omega(\vec{k})+\omega(\vec{k}-\vec{q}))}.
\end{equation}
Here, $\omega(\vec{k})=\sqrt{m^2+\vec{k}^2}$ is the energy of the
meson. We note that a summation over the relevant intermediate
states is assumed. In our calculation we use the finite range
regularisation and $u(\vec{k})$ is the ultra-violet regulator. Both
pion and kaon loops are included in the calculation. In the kaon
case, Eq.~(\ref{gea}) must of course, include the mass differences
in the intermediate states. In full QCD, the coefficients are
determined from the Lagrangian. In the quenched case, the
coefficients are obtained as in Ref.~\cite{Leinweber:2002qb} using
the quark flows of Figs.~\ref{fig:quench} and \ref{fig:quench2}. The
results are the same as those extracted within the graded symmetry
formalism. In Fig.~\ref{fig:quench2}, the diagrams with $\pi^0$ loop
are not shown since they have no contribution in the quenched or
full QCD. They do contribute to the full QCD valence sector and are
included in our calculation for the valence sector results.

The contribution to the charge form factors of Fig.~\ref{fig:diag}b
can be written as
\begin{equation}\label{geb}
G_E^{(b)}(Q^2)=\frac{\beta_E^b}{16\pi^3f_\pi^2}\int
d\vec{k}\frac{
u(\vec{k})u(\vec{k}-\vec{q})
\vec{k}\cdot (\vec{k}-\vec{q})}
{(\omega(\vec{k})+\Delta)(\omega(\vec{k}-\vec{q})+\Delta)
(\omega(\vec{k})+\omega(\vec{k}-\vec{q}))},
\end{equation}
where $\Delta$ is the positive mass difference between octet and
decuplet baryons. The contribution to the form factors of Fig.~1c is
expressed as
\begin{equation}\label{gec}
G_E^{(c)}(Q^2)=\frac{\beta_E^c}{16\pi^3f_\pi^2}\int
d\vec{k}\frac{u(\vec{k})^2}
{\omega(\vec{k}+\vec{q}/2)
+\omega(\vec{k}-\vec{q}/2)}.
\end{equation}
In the above equations, $\beta_E^i$ depends on the baryon type (or
quark type), meson loop type and quenched or full QCD in the
calculation.

In the quenched case, the double hairpin term from the
$\eta^\prime$ is expressed as
\begin{equation}
G_E^{(d)}(Q^2)=\frac{(3F-D)^2M_0^2G_E(Q^2)}{96\pi^3f_\pi^2}\int
d\vec{k}\frac{\vec{k}^2u(\vec{k})^2}
{\omega(\vec{k})^5}\equiv G_E(Q^2)G_E^d,
\label{eq:ged}
\end{equation}
where $M_0$ is the double hairpin interaction strength. As a vertex
renormalisation in the heavy-baryon limit, the $Q^2$-dependence
factorizes to define a $Q^2$-independent $G_E^d$. We should mention
that at the lowest order, the double hairpin diagram is $Q^2$
independent. The higher order terms arising from the $Q^2$
dependence of the contributions of this graph at the masses probed
in the lattice simulations are negligible.

\begin{table}[h]
\caption{Coefficient of proportionality for each diagram of
Fig.~\ref{fig:quench} for the proton.}
\begin{ruledtabular}
\begin{tabular}{ccccc}
 (a) & (b) & (c)(e)(i) & (d)(f)(h)& (g)  \\ \hline
$(D+F)^2-\frac23D^2-2F^2$ & $-(D-F)^2$ & $\frac23D^2+2F^2$ &
$(D-F)^2$ & $\frac12(D+F)^2-\frac53D^2-3F^2+2DF$
\\
\end{tabular}
\end{ruledtabular}
\end{table}

\begin{table}[h]
\caption{Coefficient of proportionality for each diagram of
Fig.~\ref{fig:quench2} for the proton.}
\begin{ruledtabular}
\begin{tabular}{cccccc}
 (a) & (b) & (c)(g) & (d)(h)& (e)(i)&(f)(j)  \\ \hline
0 & 0 & 2 & -2 & 1 & -1
\\
\end{tabular}
\end{ruledtabular}
\end{table}

\begin{table}[h]
\caption{Coefficients $\beta_E^a$ for quarks in the octet-baryons in
full and quenched QCD for Fig.~1a. The intermediate meson is $\pi$.}
\begin{ruledtabular}
\begin{tabular}{c|ccc|ccc}
 & & QQCD & & & FQCD & \\ \hline
baryon$\setminus$quark & u & d & s & u & d & s \\ \hline p &
$\frac43 D^2$ & $-\frac43 D^2$ & 0 & $(D+F)^2$ & $-(D+F)^2$ & 0 \\
\hline $\Sigma^+$ & 0 & 0 & 0 & $\begin{array}{lcr} \frac23 D^2 ~~
\Lambda\pi \\ 2F^2 ~~ \Sigma\pi ~~ \end{array}$ &
$\begin{array}{lcr}
-\frac23 D^2 ~~ \Lambda\pi \\ -2F^2 ~~ \Sigma\pi ~~ \end{array}$ & 0 \\
\hline
$\Xi^0$ & 0 & 0 & 0 & $(D-F)^2$ & $-(D-F)^2$ & 0 \\
\end{tabular}
\end{ruledtabular}
\end{table}

In the above formulas, the coefficients in quenched and full QCD can
be obtained following the methodology of
Ref.~\cite{Leinweber:2002qb}. For example, the diagram Fig.~1a is
shown in detail with quark lines in Fig.~\ref{fig:quench}. For the
pion loop, in full QCD Fig.~2a and Fig.~2c make contributions, while
in the quenched case Fig.~2a and Fig.~2b make contributions.
Therefore, we need to get the coefficient for each diagram. The
coefficient for Fig.~2c is the same as Fig.~2e which is known from
the Lagrangian, since QCD is flavor blind. By subtracting this known
coefficient from the total coefficient of full QCD, we can get the
quenched one of Fig.~2a. As an example, the coefficient of each
diagram of Fig.~\ref{fig:quench} is listed in Table I. In the same
way, the diagram of Fig.~1c can be shown in detail in
Fig.~\ref{fig:quench2}. Table II gives the coefficient of each
diagram of Fig.~\ref{fig:quench2} for the proton. We can see that
the coefficients for the first two diagrams Fig.~3a and Fig.~3b are
zero which means Fig.~1c has no contribution to the proton charge
form factor in the quenched case. In fact, for other octet baryons,
this diagram has no contribution either in the quenched case.

One can also concentrate on each quark contribution to the form
factors. All the coefficients for each quark are shown in Tables
III-VIII. In these Tables, only three baryons are listed. The
coefficients for neutron, $\Sigma^{-}$, $\Sigma^{0}$, $\Xi^{-}$ and
$\Lambda$ can be obtained by the following charge symmetry
relations:
\begin{equation}
u_n=d_p, ~~ d_n=u_p, ~~ s_n=s_p,
\label{eq:cs1}
\end{equation}
\begin{equation}
u_{\Sigma^{-}}=d_{\Sigma^{+}}, ~~ d_{\Sigma^{-}}=u_{\Sigma^{+}}, ~~
s_{\Sigma^{-}}=s_{\Sigma^{+}},
\label{eq:cs2}
\end{equation}
\begin{equation}
u_{\Xi^{-}}=d_{\Xi^{0}}, ~~ d_{\Xi^{-}}=u_{\Xi^{0}}, ~~
s_{\Xi^{-}}=s_{\Xi^{0}},
\label{eq:cs3}
\end{equation}
\begin{equation}
u_{\Sigma^{0}}=\frac12 (u_n+u_{\Xi^0}),~~ d_{\Sigma^{0}}=\frac12
(d_p+d_{\Xi^-}),~~ s_{\Sigma^{0}}=\frac12
(s_{\Sigma^+}+s_{\Sigma^-}),~~
\label{eq:cs4}
\end{equation}
\begin{equation}
u_{\Lambda}=d_{\Lambda}=\frac13\left[u_p+d_p+u_{\Xi^0}+d_{\Xi^0}
-\frac12(u_{\Sigma^+}+d_{\Sigma^-})\right],~~
s_{\Lambda}=\frac13(2s_p+2s_{\Xi^0}-s_{\Sigma^+}).
\label{eq:cs5}
\end{equation}

We express our charge form factors as
\begin{equation}\label{ge1}
G_E(Q^2)=Z-\frac16(a_0+a_2m_\pi^2+a_4m_\pi^4)Q^2+\sum_{i=a}^d
G_E^{(i)}(Q^2),
\end{equation}
where $Z$ is the wave function renormalization constant of charge
expressed as $Z=G_E(Q^2=0)-\sum_{i=a}^d G_E^{(i)}(Q^2=0)$.
$G_E(Q^2=0)$ is the charge of the baryon. $G_E^{(i)}$ is expressed
in Eqs.~(\ref{gea})-(\ref{eq:ged}). Therefore, $G_E(Q^2)$ can be
written as
\begin{eqnarray}\label{ge2}
G_E(Q^2)&=&(G_E(Q^2=0)-\sum_{i=a}^d G_E^{(i)}(Q^2=0)
-\frac16(a_0+a_2m_\pi^2+a_4m_\pi^4)Q^2+G_E^{(a)}(Q^2) \nonumber \\
&&+G_E^{(b)}(Q^2)+G_E^{(c)}(Q^2))/(1-G_E^d).
\end{eqnarray}
In anticipation of considering charge radii, we have defined the
contribution from the double-hairpin, Fig.~1d, to be proportional to
the renormalised form factor using the factorization defined in
Eq.~(\ref{eq:ged}). This follows a similar procedure to that
outlined in Ref.~\cite{Young:2004tb}.

The expansion in $a_i$ characterizes the non-chiral quark-mass
dependence of the electric charge radius of each baryon. To the
leading order we work in this manuscript, $a_0$ acts as a
counter-term to the loops and thereby removing scale-dependence of
the formal expansion. Previous works have shown that including a
partial contribution from the next higher analytic order beyond
which one is working (e.g.~Refs.~\cite{Young:2002ib,Bernard:2003rp})
can reduce the dependence on the regularisation, motivating the
$a_2$ parameter. We have also included $a_4$, which has simply been
included to better describe the lattice data, such that its
justification is purely empirical (and not mathematical). Its
presence mirrors the success obtained with a similar approach for
the octet baryon magnetic moments.  Thereby our form should only be
seen to incorporate the leading-logarithmic behavior of the EFT with
the associated counterterm.

The Sachs charge radius is defined by
\begin{equation}
<r^2>_E=-6\frac{dG_E(Q^2)}{dQ^2}|_{Q^2=0}.
\end{equation}
From the expression of $G_E(Q^2)$, $<r^2>_E$ can be written as
\begin{equation}\label{radii}
<r^2>_E=(a_0+a_2m_\pi^2+a_4m_\pi^4-6\frac{d(G_E^{(a)}(Q^2)+G_E^{(b)}(Q^2)
+G_E^{(c)}(Q^2))}{dQ^2}|_{Q^2=0})/(1-G_E^d).
\end{equation}
The above free  parameters $a_0$, $a_2$ and $a_4$ are to be
determined by fitting quenched lattice results with the described
quenched loop integrals. The octet charge radii was investigated in
Ref.~\cite{Kubis:1999xb,Kubis:2000aa} where the $SU(3)$ symmetry was
applied. Our approach is based on an $SU(2)$ framework where the
strange quark mass is held fixed, and the light quarks are always
degenerate. Thereby each baryon isospin multiplet will carry
independent LECs. Effectively there will be two (independent) $a_i$
for each multiplet. We do not impose the symmetry breaking patterns
of $SU(3)$, as we regard the strange quark mass as a scale that
cannot be described well by this low-order expansion. We note it
could be interesting to compare our effective $SU(2)$ LECs with
those extracted from earlier studies. We do not do this, but we have
now included a comparison of the predictions for the radii of the
physical states. In Ref.~\cite{Tiburzi:2008bk}, the authors have
developed two-flavor $\chi$PT to describe hyperons which are
embedded into $SU(2)$ multiplets.

\begin{sidewaystable}[h]
\caption{Coefficients $\beta_E^a$ for quarks in the octet-baryons in
full and quenched QCD for Fig.~1a. The intermediate meson is $K$.}
\begin{small}
\begin{tabular}{c|ccc|ccc}\hline\hline
 & & QQCD & & & FQCD & \\ \hline
baryon$\setminus$quark & u & d & s & u & d & s \\ \hline p & 0 & 0 &
0 & $\begin{array}{c} \frac16 (3F+D)^2 ~~ \Lambda K \\ \frac12
(D-F)^2 ~~ \Sigma K \end{array}$ & $(D-F)^2$ & $\begin{array}{c}
-\frac16 (3F+D)^2 ~~ \Lambda K \\ -\frac32 (D-F)^2 ~~ \Sigma K
\end{array}$
\\ \hline
$\Sigma^+$ & $\begin{array}{c} \frac13 D^2-F^2+2DF ~~ \Xi K \\
(D-F)^2 ~~ NK \end{array}$ & 0 & $\begin{array}{c} -(D-F)^2 ~~ NK
\\ -\frac13 D^2+F^2-2DF ~~ \Xi K \end{array}$ & $(D+F)^2$ & $-(D-F)^2$ &
$\begin{array}{c} (D-F)^2 ~~ NK \\ -(D+F)^2 ~~ \Xi K \end{array}$
\\ \hline
$\Xi^0$ & $\begin{array}{c} -\frac12 (D+F)^2+\frac16 (3F-D)^2 ~~
\Sigma K \\ -(D-F)^2 ~~ \Omega K \end{array}$ & 0 &
$\begin{array}{c} \frac12 (D+F)^2-\frac16 (3F-D)^2 ~~ \Sigma K \\
(D-F)^2 ~~ \Omega K \end{array}$ & $-(D+F)^2$ & $\begin{array}{c}
-\frac16 (3F-D)^2 ~~ \Lambda K \\ -\frac12 (D+F)^2 ~~ \Sigma K
\end{array}$
& $\begin{array}{c} \frac16 (3F-D)^2 ~~ \Lambda K \\ -\frac32 (D+F)^2 ~~ \Sigma K \end{array}$ \\
\hline\hline
\end{tabular}
\end{small}
\end{sidewaystable}

\begin{table}[h]
\caption{Coefficients $\beta_E^b$ for quarks in the octet-baryons in
full and quenched QCD for Fig.~1b. The intermediate meson is $\pi$.}
\begin{ruledtabular}
\begin{tabular}{c|ccc|ccc}
 & & QQCD & & & FQCD & \\ \hline
baryon$\setminus$quark & u & d & s & u & d & s \\ \hline p &
$-\frac{1}{3}{\cal C}^2$ & $\frac{1}{3}{\cal C}^2$ & 0 &
$-\frac{4}{9}{\cal C}^2$ & $\frac{4}{9}{\cal C}^2$ & 0 \\ \hline
$\Sigma^+$ &
0 & 0 & 0 & $\frac{1}{9}{\cal C}^2$ & $-\frac{1}{9}{\cal C}^2$ & 0 \\
\hline
$\Xi^0$ & 0 & 0 & 0 & $\frac{2}{9}{\cal C}^2$ & $-\frac{2}{9}{\cal C}^2$ & 0 \\
\end{tabular}
\end{ruledtabular}
\end{table}

\begin{table}[h]
\caption{Coefficients $\beta_E^b$ for quarks in the octet-baryons in
full and quenched QCD for Fig.~1b. The intermediate meson is $K$.}
\begin{ruledtabular}
\begin{tabular}{c|ccc|ccc}
 & & QQCD & & & FQCD & \\ \hline
baryon$\setminus$quark & u & d & s & u & d & s \\ \hline p & 0 & 0 &
0 & $\frac{1}{9}{\cal C}^2$ & $\frac{2}{9}{\cal C}^2$ &
$-\frac{1}{3}{\cal C}^2$ \\ \hline $\Sigma^+$ & $\begin{array}{c}
-\frac49 {\cal C}^2 ~~ \Delta K \\ \frac19 {\cal C}^2 ~~ \Xi^* K
\end{array}$ & 0 & $\begin{array}{c} \frac49 {\cal C}^2 ~~ \Delta K
\\
-\frac19 {\cal C}^2 ~~ \Xi^* K \end{array}$ & $\begin{array}{c}
-\frac23 {\cal C}^2 ~~ \Delta K \\ \frac29 {\cal C}^2 ~~ \Xi^* K
\end{array}$ & $-\frac{2}{9}{\cal C}^2$ & $\begin{array}{c} -\frac89 {\cal C}^2 ~~
\Delta K \\ -\frac29 {\cal C}^2 ~~ \Xi^* K \end{array}$
\\ \hline
$\Xi^0$ & $\begin{array}{c} -\frac19 {\cal C}^2 ~~ \Sigma^* K \\
\frac49 {\cal C}^2 ~~ \Omega K \end{array}$ & 0 & $\begin{array}{c}
\frac19 {\cal C}^2 ~~ \Sigma^* K \\ -\frac49 {\cal C}^2 ~~ \Omega K
\end{array}$ & $\begin{array}{c} -\frac29 {\cal C}^2 ~~ \Sigma^* K \\ \frac23
{\cal C}^2 ~~ \Omega K \end{array}$ & $-\frac{1}{9}{\cal C}^2$
& $\begin{array}{c} \frac13 {\cal C}^2 ~~ \Sigma^* K \\ -\frac23 {\cal C}^2 ~~ \Omega K \end{array}$ \\
\end{tabular}
\end{ruledtabular}
\end{table}

\begin{table}[h]
\caption{Coefficients $\beta_E^c$ for quarks in the octet-baryons in
full and quenched QCD for Fig.~1c. The intermediate meson is $\pi$.}
\begin{ruledtabular}
\begin{tabular}{c|ccc|ccc}
 & & QQCD & & & FQCD & \\ \hline
baryon$\setminus$quark & u & d & s & u & d & s \\ \hline p & 0 & 0 &
0 & 1 & -1 & 0 \\ \hline $\Sigma^+$ & 0 & 0 & 0 & 2 & -2 & 0 \\
\hline
$\Xi^0$ & 0 & 0 & 0 & 1 & -1 & 0 \\
\end{tabular}
\end{ruledtabular}
\end{table}

\begin{table}[h]
\caption{Coefficients $\beta_E^c$ for quarks in the octet-baryons in
full and quenched QCD for Fig.~1c. The intermediate meson is $K$.}
\begin{ruledtabular}
\begin{tabular}{c|ccc|ccc}
 & & QQCD & & & FQCD & \\ \hline
baryon$\setminus$quark & u & d & s & u & d & s \\ \hline
p & 0 & 0 & 0 & 2 & 1 & -3 \\ \hline
$\Sigma^+$ & 0 & 0 & 0 & 1 & -1 & 0 \\ \hline
$\Xi^0$ & 0 & 0 & 0 & -1 & -2 & 3 \\
\end{tabular}
\end{ruledtabular}
\end{table}

\section{numerical results}

In the numerical calculations, the parameters are chosen as $D=0.76$
and $F=0.50$ ($g_A=D+F=1.26$). The coupling constant ${\cal C}$ is
chosen to be $-2D$, as estimated by $SU(6)$ relations --- which
gives a similar value to that obtained from the hadronic decay width
of the $\Delta$.

Here the finite-range regulator is chosen to take the dipole form
\begin{equation}
u(k)=\frac1{(1+k^2/\Lambda^2)^2},
\end{equation}
with $\Lambda = 0.8\pm 0.1\gev$. This selected range of $\Lambda$
for the dipole has been found to give good quantitative estimates of
quenching artifacts for baryon masses \cite{Young:2002cj} and
magnetic moments \cite{Leinweber:2004tc}.

Using a fixed strange-quark mass, we estimate the $K$-meson mass to
obey the relationship
\begin{equation}
m_K^2=\frac12 m_\pi^2+m_K^2|_{\rm phy}-\frac12 m_\pi^2|_{\rm phy}\,.
\label{eq:mK}
\end{equation}

\begin{center}
\begin{figure}[hbt]
\includegraphics[angle=90,scale=0.45]{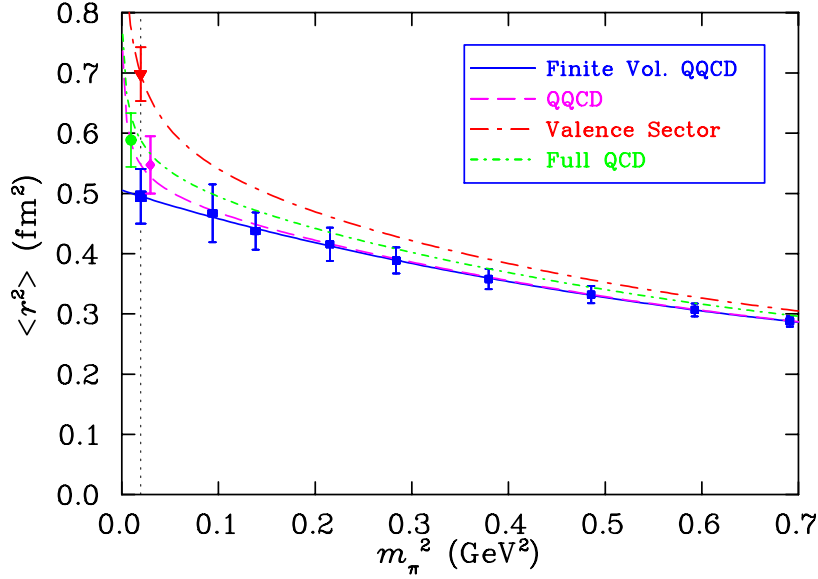}
\caption{The contribution of a single $u$ quark with unit charge to
the proton charge radius versus pion mass. The square, rhombus,
triangle, and round symbols are for the finite volume quenched QCD,
infinite volume quenched QCD, valence sector and full QCD results,
respectively. \label{fig:u-Nucleon}}
\end{figure}
\end{center}

\begin{center}
\begin{figure}[hbt]
\includegraphics[angle=90,scale=0.45]{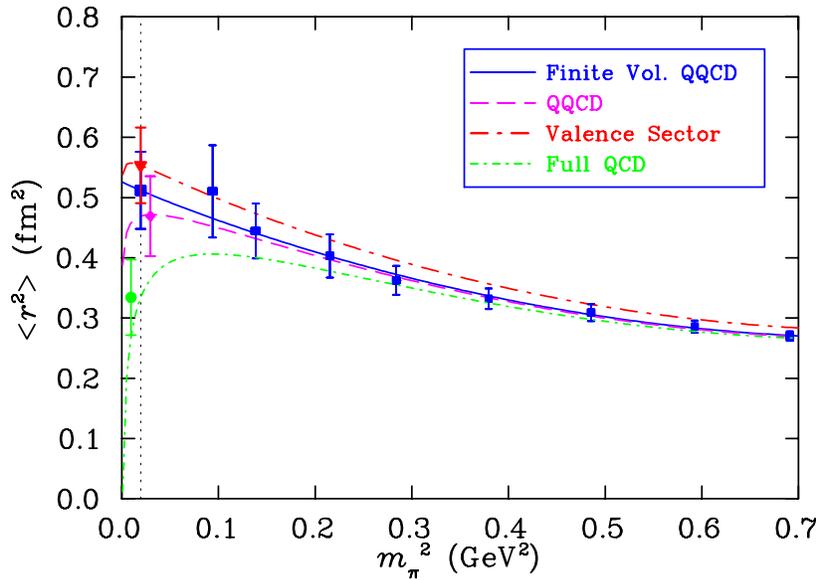}
\caption{The contribution of a $d$ quark with unit charge to the
proton charge radius versus pion mass. The square, rhombus,
triangle, and round symbols are for the finite volume quenched QCD,
infinite volume quenched QCD, valence sector and full QCD results,
respectively. \label{fig:d-Nucleon}}
\end{figure}
\end{center}

We first study the $u$-quark contribution to the proton charge
radius. Four kinds of extrapolations are shown in
Fig.~\ref{fig:u-Nucleon}. The square, rhombus, triangle, and round
symbols are for the finite volume quenched QCD, infinite volume
quenched QCD, valence sector and full QCD results, respectively. We
remind the reader that the ``valence" result denotes the connected
current insertions in Full QCD. The quenched lattice results are
described very well. For infinite volume, all of them have
log-divergent behavior at $m_\pi=0$, which means that the pion cloud
extends to infinity for a massless pion. For finite volume, the
integration is replaced by the summation of the momentum which shows
no log divergence in the chiral limit. Technically, one should be
cautious when considering the finite volume curve in the domain
where $\mpi L$ becomes small. Ideally, one would like to stay in the
regime of $m_\pi L>6$ (or $2\pi$). Ambitious lattice calculations
push this down to 4 or 3, at which point one may be approaching the
limits of describing the finite-volume corrections by the one-loop
EFT.  This limit, for our 2.56 fm box, is at pion masses of order
250 to 300 MeV, the lightest pion mass considered in the lattice
simulations. This is not of serious concern, as {\em corrections}
are performed at finite volume and {\em extrapolations} subsequently
performed at infinite volume.

Fig.~\ref{fig:d-Nucleon} displays the $d$-quark contribution to the
proton charge radius. The lines with different types have the same
meaning as in Fig.~4.  For infinite volume, the charge radius bends
down as $m_\pi$ approaches zero, indicating that the $\bar{d}$
contributes more strongly to the long-ranging tail than the $d$ in
accord with the predominant $p\rightarrow n\pi$ channel. At
$m_\pi=0$, the curves also have log divergent behavior. We note that
the positive charge radius is much like a constituent-quark type
expectation, and that the unusual chiral features are only
anticipated to become dominant far below the physical quark mass.

\begin{center}
\begin{figure}[hbt]
\includegraphics[angle=90,scale=0.45]{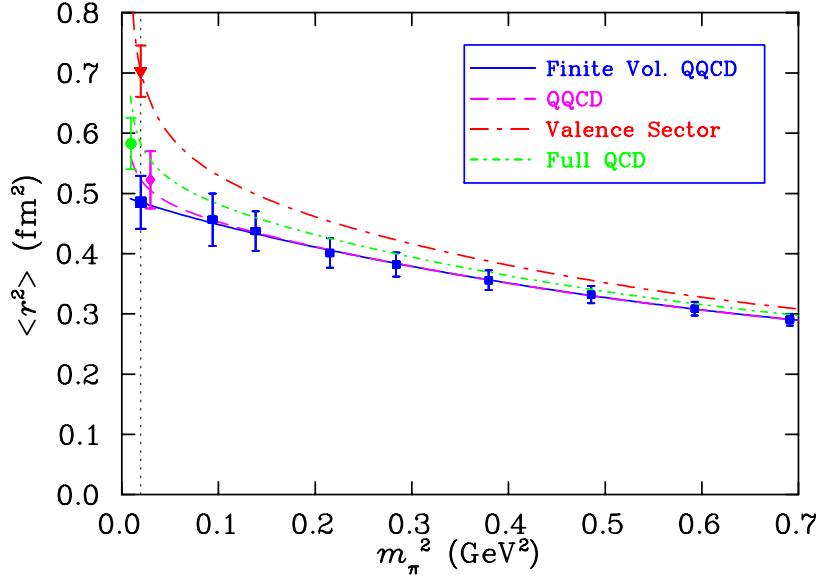}
\caption{The contribution of a single $u$ quark with unit charge to
the $\Sigma^+$ charge radius versus pion mass. The square, rhombus,
triangle, and round symbols are for the finite volume quenched QCD,
infinite volume quenched QCD, valence sector and full QCD results,
respectively. \label{fig:u-Sig}}
\end{figure}
\end{center}

\begin{center}
\begin{figure}[hbt]
\includegraphics[angle=90,scale=0.45]{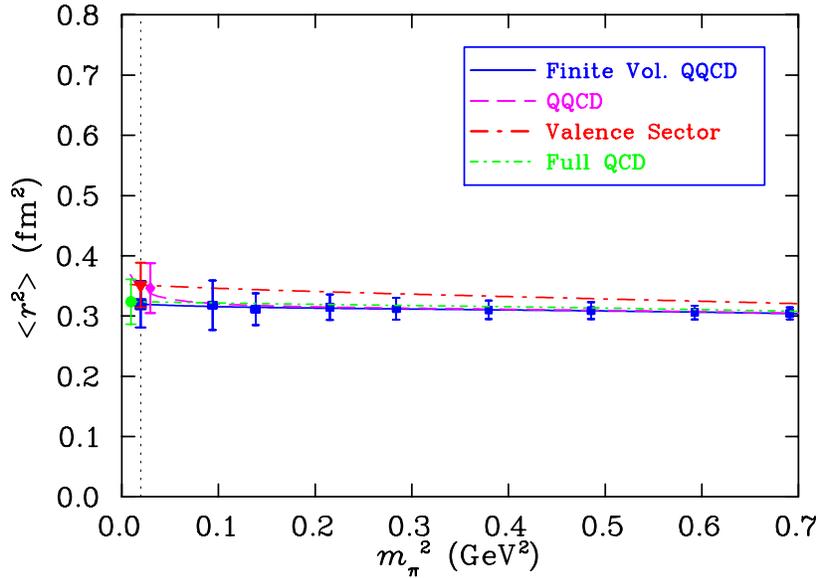}
\caption{The contribution of an $s$ quark with unit charge to the
$\Sigma^+$ charge radius versus pion mass. The square, rhombus,
triangle, and round symbols are for the finite volume quenched QCD,
infinite volume quenched QCD, valence sector and full QCD results,
respectively.
\label{fig:s-Sig}}
\end{figure}
\end{center}

We now discuss the quark distribution radius in strange octet
baryons. The $u$-quark contribution to the $\Sigma^+$ charge radius
is shown in Fig.~\ref{fig:u-Sig}. In contrast to the proton case, in
quenched QCD there is no pion loop contribution to the $\Sigma^+$
charge radius. The non trivial loop contributions involve the $K$
meson. Consequently, the $u$-quark distribution radius has no log
divergence (as the $SU(2)$ chiral limit is approached). In full QCD,
both the valence and total sectors exhibit log-divergent charge
radii.  At the physical pion mass, the total $u$-quark distribution
radius is found to be very similar to that in the proton.

The singly-represented quark of the $\Sigma^+$ is a strange quark.
We show its contribution to the $\Sigma^+$ charge radius in
Fig.~\ref{fig:s-Sig}.  Since there cannot be any leading order
pion-loop contributions to the $s$-quark radius the chiral
corrections are much less dramatic than in the light-quark sector.
One may also compare $K$ contributions at the physical point with
those of the light-quark sector as observed near $\mpi^2\sim
m_{K({\rm phys})}^2\sim 0.25\gev^2$. Because the $s$ quark mass is
held fixed, any variation is due to an environment effect associated
with the light quarks.

\begin{center}
\begin{figure}[hbt]
\includegraphics[angle=90,scale=0.45]{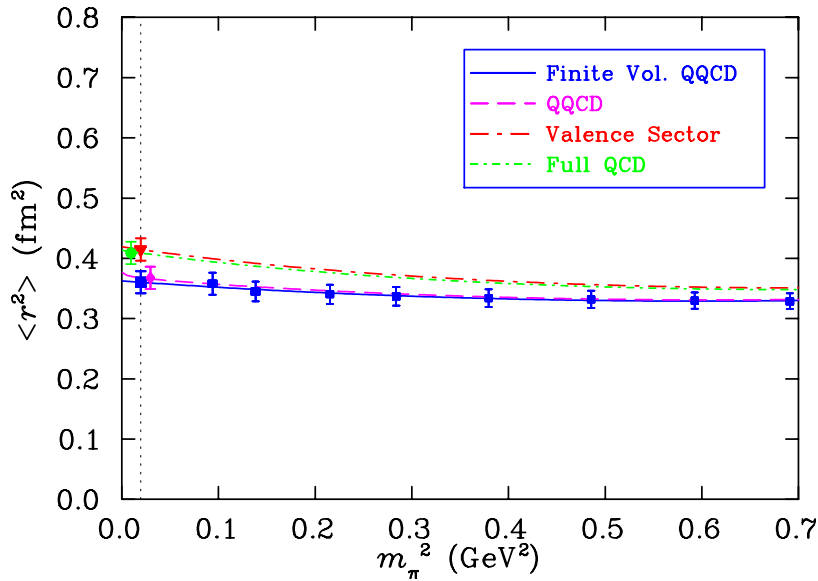}
\caption{The contribution of a single $s$ quark with unit charge to
the $\Xi^0$ charge radius versus pion mass. The square, rhombus,
triangle, and round symbols are for the finite volume quenched QCD,
infinite volume quenched QCD, valence sector and full QCD results,
respectively.
\label{fig:s-Xi}}
\end{figure}
\end{center}

\begin{center}
\begin{figure}[hbt]
\includegraphics[angle=90,scale=0.45]{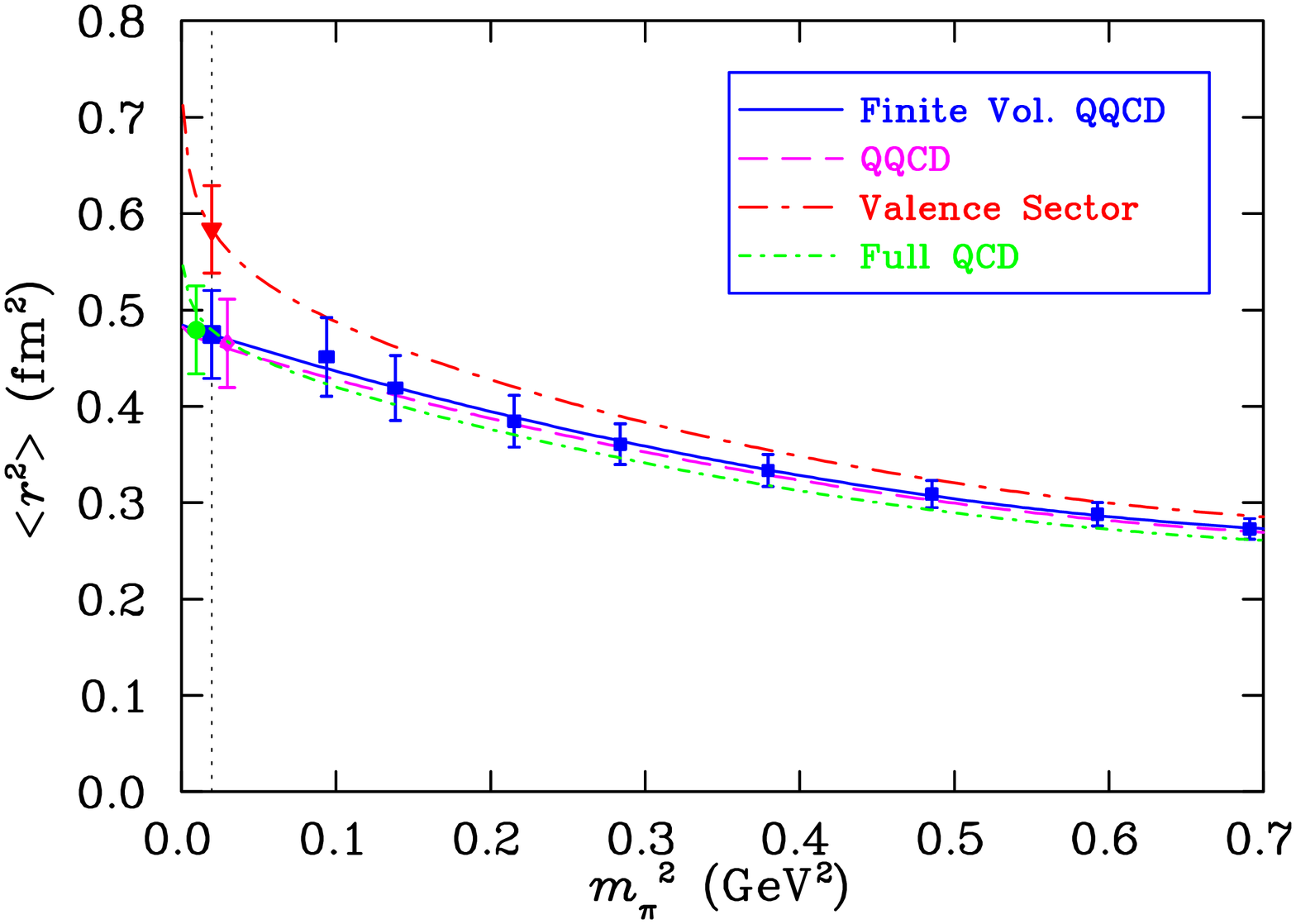}
\caption{The contribution of a $u$ quark with unit charge to the
$\Xi^0$ charge radius versus pion mass. The square, rhombus,
triangle, and round symbols are for the finite volume quenched QCD,
infinite volume quenched QCD, valence sector and full QCD results,
respectively.
\label{fig:u-Xi} }
\end{figure}
\end{center}

\begin{center}
\begin{figure}[hbt]
\includegraphics[angle=90,scale=0.45]{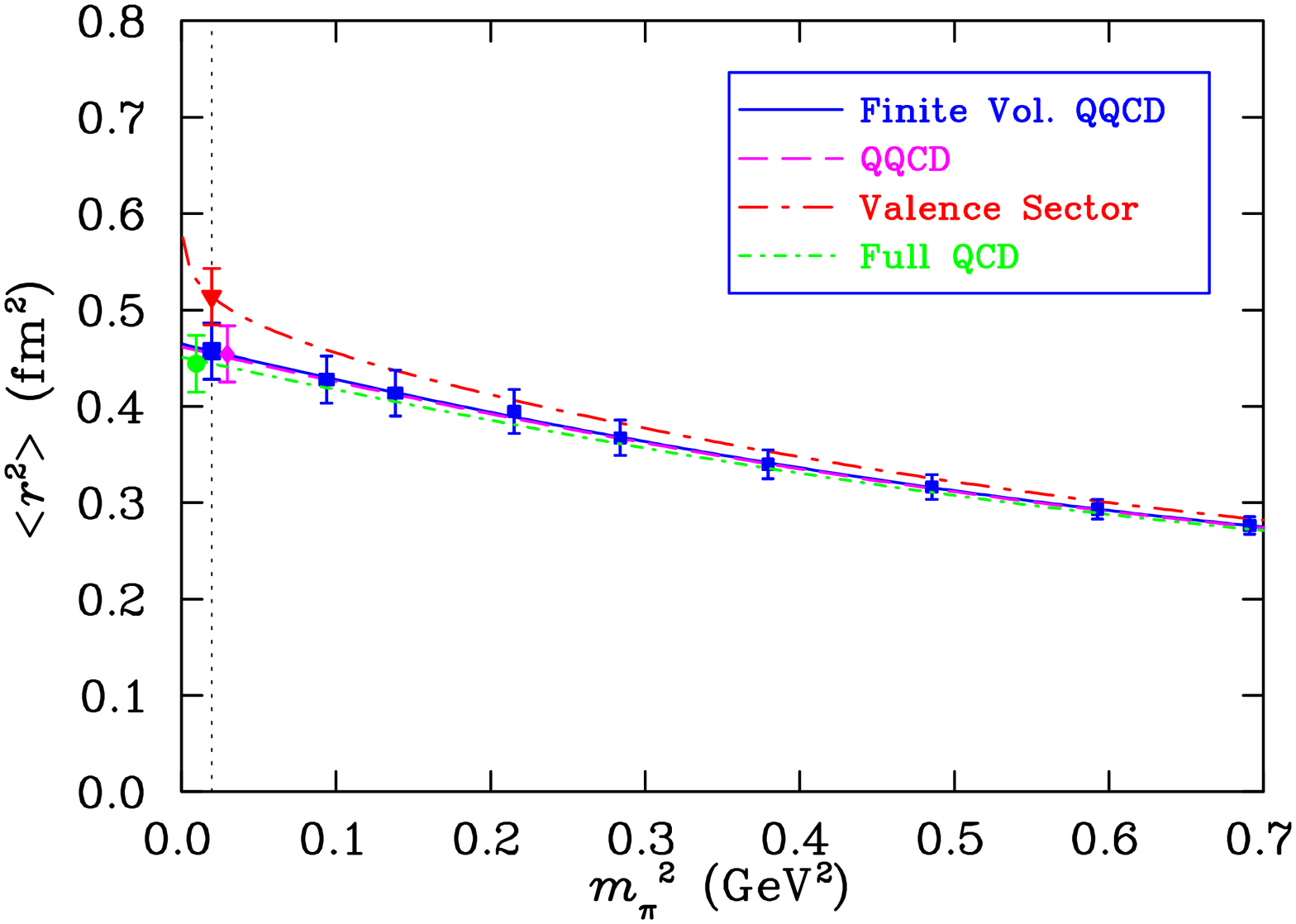}
\caption{The contribution of a single light quark with unit charge
to the $\Lambda$ charge radius versus pion mass. The square,
rhombus, triangle, and round symbols are for the finite volume
quenched QCD, infinite volume quenched QCD, valence sector and full
QCD results, respectively.
\label{fig:u-Lambda}}
\end{figure}
\end{center}

\begin{center}
\begin{figure}[hbt]
\includegraphics[angle=90,scale=0.45]{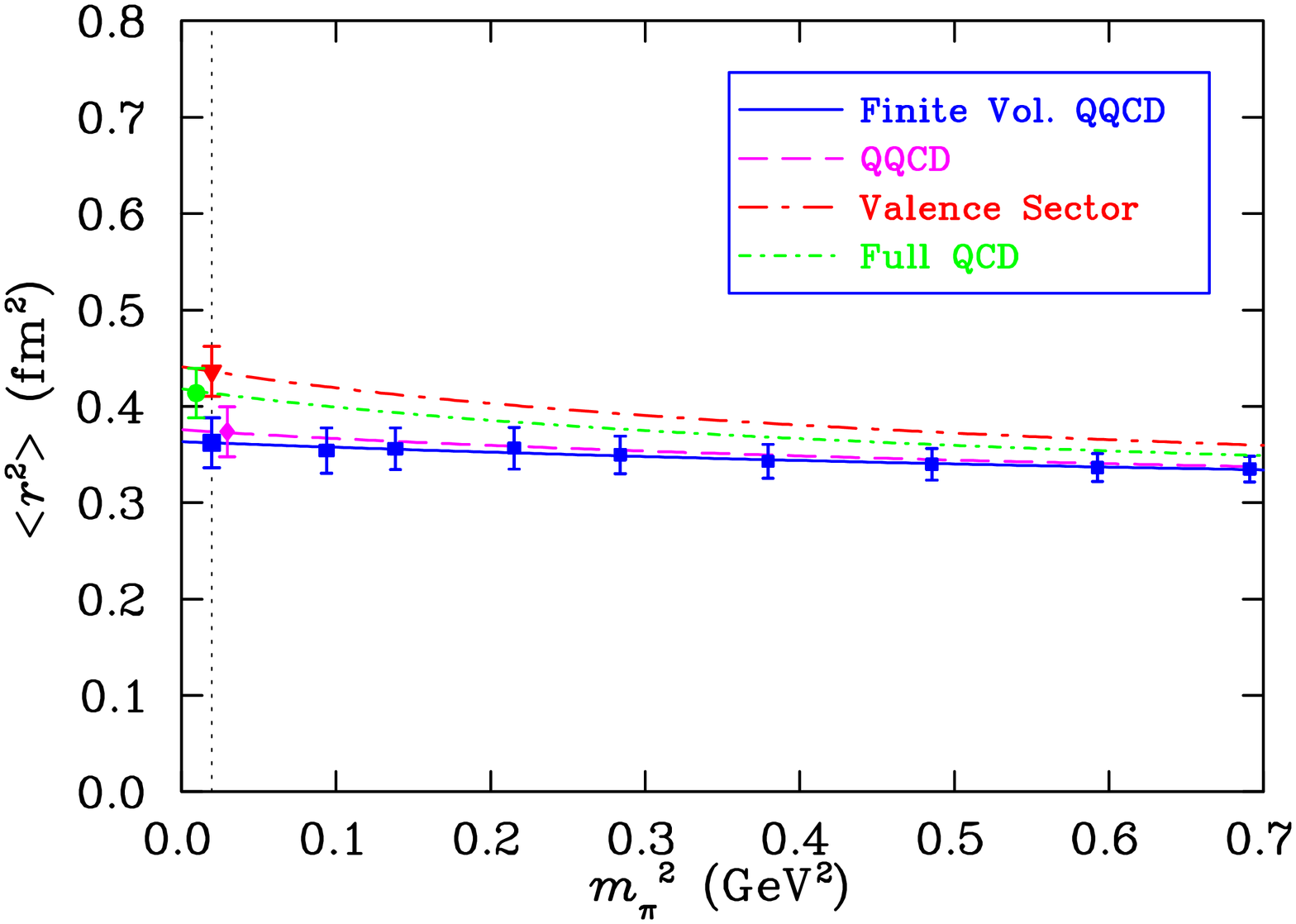}
\caption{The contribution of an s quark with unit charge to the
$\Lambda$ charge radius versus pion mass. The square, rhombus,
triangle, and round symbols are for the finite volume quenched QCD,
infinite volume quenched QCD, valence sector and full QCD results,
respectively.
\label{fig:s-Lambda}}
\end{figure}
\end{center}

\begin{center}
\begin{figure}[hbt]
\includegraphics[angle=90,scale=0.45]{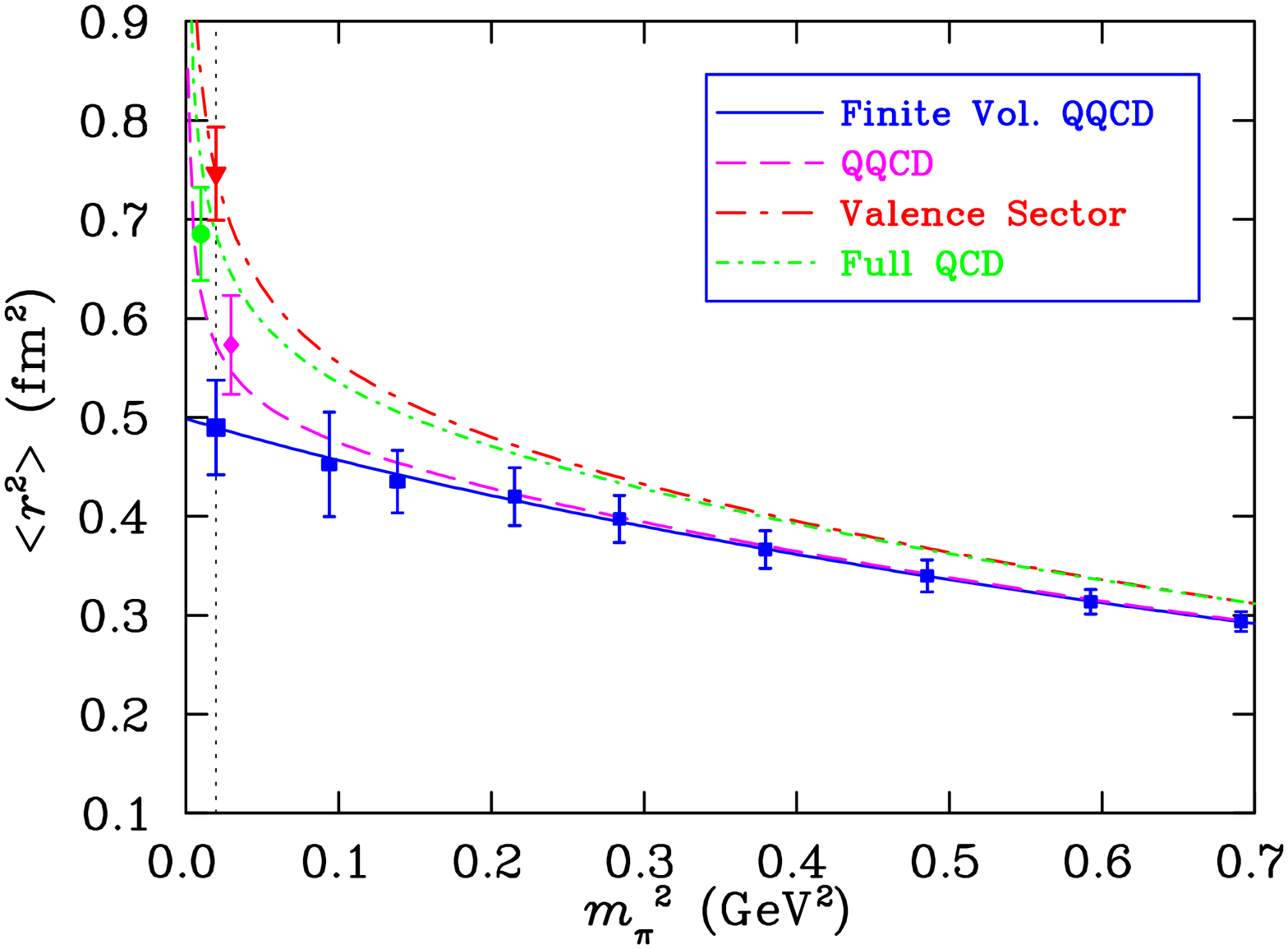}
\caption{The proton charge radius versus pion mass. The square,
rhombus, triangle, and round symbols are for the finite volume
quenched QCD, infinite volume quenched QCD, valence sector and full
QCD results, respectively. \label{fig:proton}}
\end{figure}
\end{center}

The $\Xi^0$ is composed of two strange quarks and one up quark. The
strange quark contribution to $\Xi^0$ is shown in
Fig.~\ref{fig:s-Xi}. Similar to the $s$ quark in the $\Sigma^+$, all
the curves display a very mild environment dependence on the
light-quark mass.

The $u$-quark contribution to the $\Xi^0$ is shown in
Fig.~\ref{fig:u-Xi}. In full QCD, both the full and valence
contributions are divergent as $m_\pi\to 0$. Here we see that the
charge radius diverges in the positive direction, opposite to that
of the $d$ quark in the proton. Here, the pion loop must contain a
light anti-quark from the sea and thereby the valence $u$ is pure
quark (that is, it cannot be an anti-quark, or it cannot couple to a
pion through a $Z$-type diagram).

The single light quark and strange quark contribution to the
$\Lambda$ is shown in Fig.~\ref{fig:u-Lambda} and
\ref{fig:s-Lambda}, respectively. It is interesting that the light
quark sector does not show much curvature. This is consistent with
the chiral coefficients for the leading-order pion dressing
vanishing. The contribution from the $\pi^+$ and $\pi^-$ meson
clouds cancel each other. The s quark in the $\Lambda$ couples
strongly to $NK$ and it looks like there is some environment
dependence.

Using the charge-symmetry relations above,
Eqs.~(\ref{eq:cs1})--(\ref{eq:cs5}), one can reconstruct any desired
baryon form factors by applying the appropriate charge factors.

We now discuss the constructed octet-baryon charge radii. The proton
charge radii are illustrated in Fig.~\ref{fig:proton}. At the
physical pion mass, we find the proton charge radius is about
$0.69\pm 0.05\fm^2$. The $u$, $d$ and $s$ quarks all provide a
contribution to the proton charge radii. We note that the valence
sector alone contributes a large fraction to the total charge
radius.  There is a small correction by adding all the disconnected
contributions of the 3 light-quark flavors in the process of
correcting QQCD via FRR EFT. Further, the strangeness component only
constitutes a small part of this small correction
\cite{Leinweber:2006ug}.

\begin{center}
\begin{figure}[hbt]
\includegraphics[angle=90,scale=0.45]{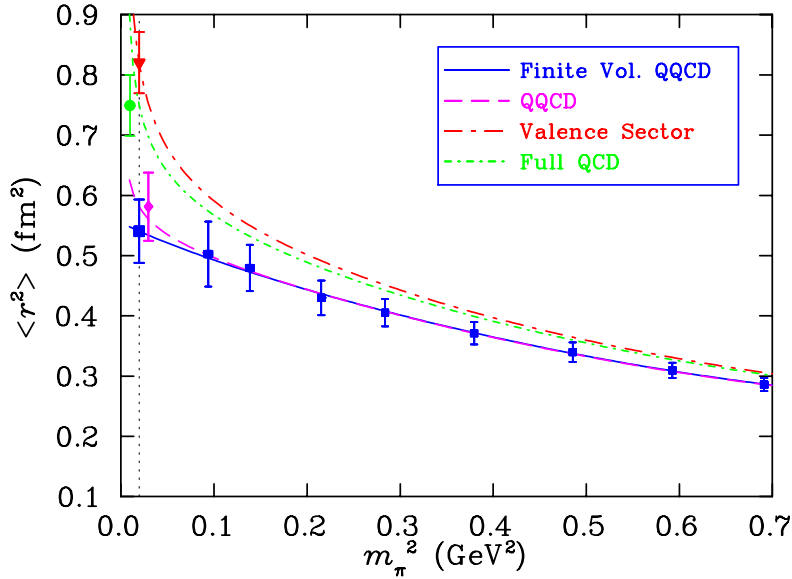}
\caption{$\Sigma^+$ charge radius versus pion mass. The square,
rhombus, triangle, and round symbols are for the finite volume
quenched QCD, infinite volume quenched QCD, valence sector and full
QCD results, respectively.\label{fig:Sig}}
\end{figure}
\end{center}

The charge radius of $\Sigma^+$ is shown in Fig.~\ref{fig:Sig}. The
numerical value is observed to be a little larger than that of
proton, with the physical charge radius of the $\Sigma^+$ being
$0.75\pm 0.05\fm^2$. The primary cause of this radius being larger
than the proton is the fact that the (negatively-charged) strange
quark in the $\Sigma$ has a narrower spatial distribution than the
down quark in the proton.

\begin{center}
\begin{figure}[hbt]
\includegraphics[angle=90,scale=0.45]{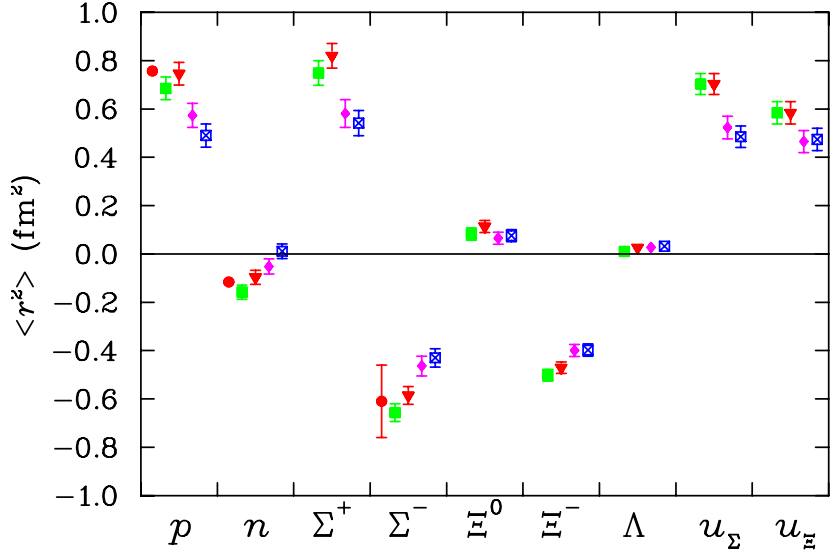}
\caption{Octet-baryon charge radii at the physical pion mass. The
hollow square, solid rhombus, triangle and square symbols are for
the finite volume quenched QCD, infinite volume quenched QCD,
valence sector and full QCD results, respectively. The experimental
data for proton, neutron and $\Sigma^-$ is shown with the left-most
bullet.\label{fig:all}}
\end{figure}
\end{center}

In Fig.~\ref{fig:all}, the charge radii of the octet baryons at the
physical pion mass are shown. The extrapolated physical radii of the
proton, neutron and $\Sigma^-$ are in good agreement with the
experimental data. The radius of $\Xi^0$ is positive, in contrast to
the neutron radius. The reason is that, for the neutron, in full
QCD, the $\pi^-$ cloud is the dominant contribution, whereas for the
$\Xi^0$, the pion intermediate state is a $\pi^+$ (with a $\Xi^-$
intermediate baryon). Since both $\pi^+$ and $\pi^-$ contribute to
the $\Lambda$, the radius of the $\Lambda$ is close to zero. This is
perhaps surprising as one would expect the net positive charge of
the light quarks to dominate more.

The results of our extrapolation and unquenching are summarized in
Table~\ref{tab:res}. Here, we see a careful breakdown of the
propagation of uncertainties at the various stages of calculation.

\begin{table}
  \caption{Summary of extrapolation and unquenching results. Each major
    column consists of our best value (in $\mathrm{fm}^2$), followed by
    the three dominant sources of uncertainty: statistical, lattice
    scale determination $a=0.128\pm0.006\fm$ and regularization scale
    $\Lambda=0.8\pm0.1\gev$, respectively, all quoted relative to the final digit of
    the best value. The sign of the uncertainty reflects the
    correlation with $a$ or $\Lambda$. \label{tab:res}}
\begin{ruledtabular}
\begin{tabular}{l|rrrr|rrrr|rrrr|rl}
           & \multicolumn{4}{c|}{QQCD} & \multicolumn{4}{c|}{Valence QCD} & \multicolumn{4}{c|}{Total} & \multicolumn{2}{c}{Experiment} \\
\hline
$p$        & $ 0.573$ & 50 & 45 & 8   & $ 0.746$ & 47 & 42 & 27 & $ 0.685$ & 47 & 42 & 21  & $ 0.766  \pm 0.012$  & \cite{Amsler:2008zz} \\
$n$        & $-0.052$ & 31 &  2 &-5   & $-0.097$ & 29 &  3 & -9 & $-0.158$ & 29 &  3 &-15  & $-0.1161 \pm 0.0022$ & \cite{Amsler:2008zz}\\
$\Sigma^+$ & $ 0.581$ & 57 & 53 & 6   & $ 0.820$ & 51 & 47 & 28 & $ 0.749$ & 51 & 47 & 21  \\
$\Sigma^-$ & $-0.464$ & 41 &-42 &-5   & $-0.586$ & 37 &-38 &-17 & $-0.657$ & 37 &-38 &-24  & $-0.61 \pm 0.15 $ & \cite{GoughEschrich:2001ji} \\
$\Xi^0$    & $ 0.065$ & 25 &  8 &-1   & $ 0.113$ & 25 &  8 &  4 & $ 0.082$ & 25 &  8 & -2  \\
$\Xi^-$    & $-0.400$ & 25 &-37 & 0   & $-0.471$ & 24 &-37 &-12 & $-0.502$ & 24 &-37 &-17  \\
$\Lambda$  & $ 0.027$ &  8 &  3 & 0   & $ 0.026$ &  8 &  3 & -2 & $ 0.010$ &  8 &  3 & -4  \\
$u_\Sigma$  & $ 0.523$ & 47 & 47 & 6   & $ 0.703$ & 43 & 42 & 22 & $ 0.703$ & 43 & 42 & 22  \\
$u_{\Xi}$   & $ 0.465$ & 46 & 45 &-1   & $ 0.584$ & 46 & 45 & 16 & $ 0.584$ & 46 & 45 & 16
\end{tabular}
\end{ruledtabular}
\end{table}

\section{summary}

We extrapolated state-of-the-art lattice results for the
quark-sector decomposition of the octet-baryon charge radii in
quenched heavy baryon chiral perturbation theory using FRR.  All
leading-loop diagrams have been incorporated, including all
contributions from both octet and decuplet baryon intermediate
states. Finite-range regularisation has been utilized in the one
loop calculation to improve the convergence at moderate quark
masses. Further, the use of FRR provides a separation of scales,
which has enabled the use of a demonstrated technique to obtain
estimates of the full-QCD results from the quenched lattice
simulations. We acknowledge the phenomenological aspects of our
calculation. As such, we have an empirically motivated extrapolation
form which is equivalent to the leading one-loop HBChPT form in the
domain where higher-order terms are genuinely negligible.

The individual quark contributions to the baryon charge radii in
quenched QCD have been extrapolated, with subsequent predictions for
the corresponding radii in both valence and full QCD. We note that
the valence predictions will be readily confronted with the next
generation of lattice simulations of full QCD, as only connected
current insertions are required. The contribution from pion loops is
observed to generate significantly more enhancement of the radii
than that from K mesons. Further, disconnected contributions to the
radii are rather small compared with the dominant valence
contributions.  One of the consequences of this is that the
strange-quark disconnected contributions only represent a small
component of an already small correction.

The charge radii of the proton, neutron and $\Sigma^-$ are in good
agreement with the experimental results. Our result for $\Sigma^-$
is quite accurate relative to the current experimental measurement.
Further, we have presented predictions for the remaining five
baryons, for which no experimental information exists to date. These
predictions remain to be tested by potential future measurements and
future studies in {\em ab initio} lattice QCD.

\section*{Acknowledgements}

We thank the Australian Partnership for Advanced Computing (APAC)
and eResearch South Australia for supercomputer support enabling
this project.  This work is supported by the Australian Research
Council and by U.S. DOE Contract No. DE-AC05-06OR23177, under which
Jefferson Science Associates, LLC operates Jefferson Laboratory, and
DE-AC02-06CH11357, under which UChicago Argonne, LLC operates
Argonne National Laboratory.

\end{document}